\documentclass[aps,pra,raggedbottom,reprint,amsmath,amssymb,superscriptaddress,longbibliography]{revtex4-2}

\usepackage{graphicx}
\usepackage{booktabs}
\usepackage{xcolor}
\usepackage{bm}
\usepackage{amsthm}
\usepackage[hidelinks]{hyperref}
\graphicspath{{../figures/}}

\newtheorem{theorem}{Theorem}

\newtheorem{lemma}[theorem]{Lemma}

\newcommand{\D}{\mathcal{D}}
\newcommand{\Tr}{\operatorname{Tr}}
\newcommand{\E}{\mathbb{E}}
\newcommand{\Var}{\operatorname{Var}}
\newcommand{\ssstate}{\mathrm{ss}}
\newcommand{\cont}{\mathrm{cont}}
\begin{document}

\title{Hidden collision statistics in bosonic heat transport:\\
Superthermal correlations at fixed mean current}

\author{Iu. Nosal}
\email{nosal.ia16@physics.msu.ru}
\affiliation{Department of Mathematical Physics, Steklov Mathematical Institute of Russian Academy of Sciences, ul. Gubkina 8, Moscow, 119991 Russia}
\author{A. Teretenkov}
\email{taemsu@mail.ru}
\affiliation{Department of Mathematical Physics, Steklov Mathematical Institute of Russian Academy of Sciences, ul. Gubkina 8, Moscow, 119991 Russia}

\date{September 20, 2026}

\begin{abstract}
Reservoirs with the same mean relaxation rate can be indistinguishable at the
level of average energy transport while producing different fluctuations. We
study a bosonic mode coupled to hot and cold Poisson streams of thermal
ancillas through finite beam-splitter collisions. The averaged evolution is a compound-Poisson semigroup generated by finite Gaussian event channels. Its nonequilibrium steady
state is an exact mixture of thermal states governed by a random affine fixed
point. Consequently, the mean-occupation dynamics and bath-resolved mean heat currents coincide with those
of the matched continuous Lindblad reservoir, whereas higher correlations
retain the collision strength. For equal collision transmissivities, we derive
an exact superthermal bunching law controlled by the temperature contrast and
collision strength. We also obtain the asymptotic rates of the first two heat cumulants for an ideal stationary event-resolved two-point-measurement record and
separate local one-collision contributions from temporal correlations. Fock-space diagonalization and Monte Carlo trajectories validate
the formulas and connect full resets to the weak-collision Gaussian limit.
\end{abstract}

\maketitle

\section{Introduction: mean transport can hide reservoir granularity}
\label{sec:introduction}

Collision models replace a macroscopic environment by a stream of auxiliary systems that interact with the system of interest one at a time. They provide a microscopic route to quantum channels, master equations, and stochastic thermodynamics in settings including micromasers, reservoir engineering, quantum trajectories, and transport \cite{Ciccarello2022,Strasberg2017}. Early collision-model constructions already used successive system--ancilla interactions to describe thermalization and Markovian open-system dynamics \cite{Scarani2002,Ziman2005}. A microscopic route to quantum Poisson collision dynamics is provided by the low-density limit of scattering from a dilute, not necessarily equilibrium, Bose gas \cite{Pechen2004}; this motivates Poissonian event timing but does not by itself yield the thermal beam-splitter channel used below. The interaction strength of a single collision and the rate at which collisions occur are independent physical parameters. This separation is lost in a continuous weak-coupling description, where the event rate and event strength enter only through their combined effective relaxation rate. Continuous limits of repeated quantum interactions under suitable scalings of the interaction time and coupling strength have been developed in \cite{AttalPautrat2006,AttalJoye2007}. A broader mathematical treatment of repeated-interaction systems is given in \cite{Bruneau2014}. Recent microscopic derivations have also quantified the approximation errors underlying Markovian and non-Markovian collision-model descriptions \cite{Lacroix2025}.

The Gaussian sector is particularly transparent. Beam-splitter mixing of a bosonic system and ancilla realizes a thermal attenuator channel \cite{Weedbrook2012}, and rapid Gaussian ancillary bombardment can reproduce thermalization of a harmonic mode \cite{Grimmer2018}. Random collision times have also been used as controllable noise in transport networks \cite{Chisholm2021}, while random Gaussian operations generate dissipative evolutions that preserve convex mixtures of Gaussian states \cite{Linowski2022}. More microscopic Poisson-bath constructions can produce white non-Gaussian noise and coupling-dependent saturation effects \cite{Funo2024}. These results establish that collision timing and finite event strength are physically meaningful, but they do not by themselves determine which steady transport observables retain this information.

The question is especially sharp for a mode coupled to hot and cold streams.
By matching, for each stream, the effective occupation-relaxation rate to the
corresponding thermal Lindblad rate, while keeping the same bath occupation,
the collision and continuous descriptions have identical mean-occupation
dynamics and bath-resolved mean heat currents.

Several neighboring frameworks address related but distinct questions.
Stochastic collision models have been used to study noise-assisted transport
in finite quantum networks \cite{Chisholm2021}, whereas here the transported
quantity is heat between thermal bosonic streams. Repeated-interaction
reservoirs have also been used to construct nonequilibrium steady states and
stationary currents in boundary-driven quantum systems
\cite{Karevski2009}.

Damping and correlation functions of coupled quantum oscillators interacting
with thermal reservoirs at different temperatures were studied by Glauber
and Man'ko \cite{GlauberManko1984}.

Collisional thermometry has been developed for sequential quantum probes \cite{Seah2019} and, more recently, for Gaussian systems \cite{Alves2024}. Repeated state-update models in which a temperature measurement prepares a Gibbs posterior provide a related thermodynamic measurement framework \cite{GerasimovTeretenkov2025}; unlike the number-resolved TPM protocol used here, that construction addresses thermometry rather than event-resolved heat counting. Here we instead diagnose stationary-reservoir granularity through steady-state system correlations.

Full counting statistics of quantum heat transport has been developed for
harmonic conductors \cite{SaitoDhar2007,Agarwalla2012}, while continuous
bosonic-cavity counting provides photon and heat statistics for infinitesimal
thermal damping \cite{Brange2019}; our event maps have finite strength and the
continuous model is recovered only in the joint weak-event high-rate limit.

Multibath qubit collision models describe nonequilibrium steady states and TPM heat trajectories, including settings with inter-ancilla memory \cite{McElvogue2026}. More generally, correlated and structured ancilla environments admit exact tensor-network descriptions and Markovian embeddings \cite{FilippovLuchnikov2022}. Our model instead uses independent Markovian streams and an infinite bosonic ladder. A diagonal qubit model may also reduce to a scalar population recursion; what is specifically bosonic here is that attenuation preserves the full one-parameter thermal family and converts moments of the scalar affine variable into all normally ordered field correlations. Repeated interactions of a quantized cavity field with atomic beams provide another closely related infinite-dimensional bosonic setting \cite{BruneauPillet2009}.

Against this background, the open questions are whether finite event strengths
leave exact stationary signatures beyond the matched mean energy dynamics and
whether the zero-frequency heat noise admits a closed finite-collision
expression.

Random repeated-interaction systems and their long-time asymptotics have been studied in a general mathematical setting \cite{Bruneau2008}. Earlier exact closure results for bosonic Poisson jump dynamics were obtained for quadratic unitary transformations \cite{Teretenkov2020Poisson}, and the compound-Poisson construction with multiple event types and moment closure for Poissonian dynamics generated by quadratic system--ancilla interactions are established tools \cite{Nosal2022}. The event maps here are instead nonunitary thermal attenuator channels obtained after tracing out fresh ancillas. Gaussian thermal-attenuator channels are standard in Gaussian quantum information \cite{Weedbrook2012}, while the general theory of contractive random affine recursions and perpetuities is classical \cite{Vervaat1979,Diaconis1999}. The channel-level moment closure and the algebraic moment recursion following from the affine fixed-point equation are therefore specializations of known general results. The new model-specific results are the exact first-moment transport degeneracy with the matched continuous bath, the identification of the invariant state as a collision-history-dependent thermal mixture, the explicit and invertible bunching law in the common-transmissivity family, and the closed finite-collision TPM heat-noise formula including the temporal-correlation contribution. The last result is derived directly from moment and Poisson equations for the stationary population process, without assuming the existence of a scaled cumulant generating function (SCGF) at finite counting field in the infinite-dimensional state space. We do \emph{not} obtain a full SCGF or a finite-collision fluctuation symmetry.
\section{Two thermal streams and a matched continuous reservoir}
\label{sec:model}

\begin{figure*}[t]
 \includegraphics[width=0.96\textwidth]{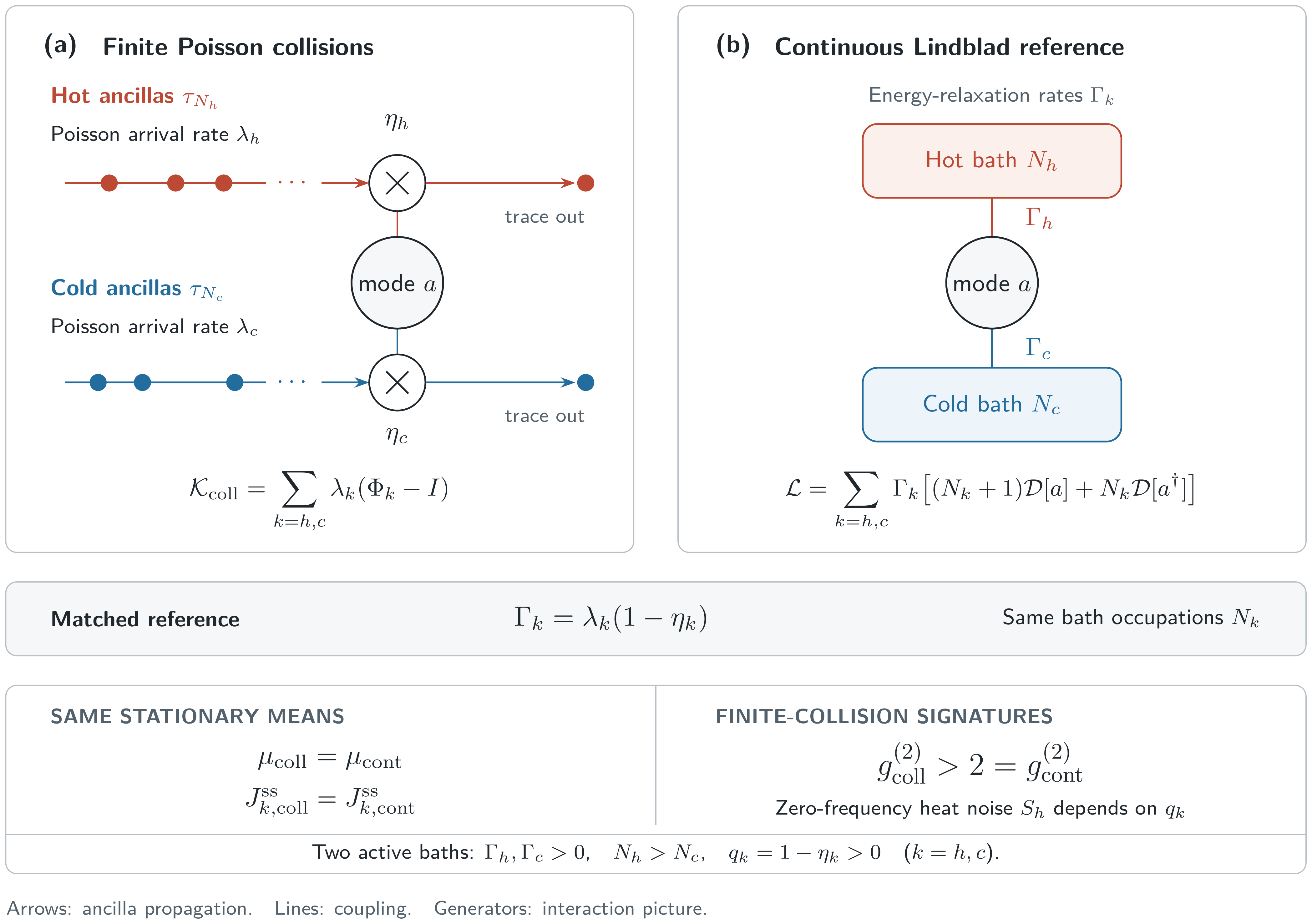}
 \caption{Finite Poisson collision reservoirs and their matched continuous
Lindblad reference.
(a) Independent Poisson streams supply fresh thermal ancillas in states
$\tau_{N_k}$ with mean occupations $N_k$ at arrival rates $\lambda_k$,
$k\in\{h,c\}$. Each ancilla interacts once with mode $a$ through a
beam-splitter unitary with system transmissivity $\eta_k$ and is then traced
out. Circles marked with a cross denote the beam-splitter interactions.
(b) The continuous reference uses the same bath occupations and
energy-relaxation rates $\Gamma_k=\lambda_k(1-\eta_k)$.
For equal initial mean occupations, the models have identical $\mu(t)$ and
bath-resolved mean heat currents $J_k(t)$; their stationary mean occupations
and currents also coincide, with $J_h^{\mathrm{ss}}=-J_c^{\mathrm{ss}}$.
For two active reservoirs with $N_h>N_c$ and finite collision strengths
$q_k=1-\eta_k>0$, the collision model exhibits
$g_{\mathrm{coll}}^{(2)}>2=g_{\mathrm{cont}}^{(2)}$.
Stationary higher-order correlations $g^{(m)}$, $m\geq2$, and zero-frequency
heat noise $S_h$ depend on the collision strengths at fixed $\Gamma_k$.
Horizontal arrows indicate ancilla propagation; the other connections
indicate couplings. The displayed generators are given in the interaction
picture.}
 \label{fig:schematic}
\end{figure*}

\subsection{Energy-preserving Gaussian collisions}
The system is a single bosonic mode with annihilation operator $a$, number operator
$n=a^\dagger a$, and Hamiltonian
\begin{equation}
 H_S=\hbar\omega n.
 \label{eq:HS}
\end{equation}
Ancillas of type $k\in\{h,c\}$ are independent bosonic modes $b_k$ prepared in
thermal states
\begin{align}
 \tau_{N_k}&=\frac{1}{N_k+1}\sum_{r=0}^{\infty}
 \left(\frac{N_k}{N_k+1}\right)^r |r\rangle\langle r|,
 \nonumber\\
 N_k&=\frac{1}{e^{\beta_k\hbar\omega}-1}.
 \label{eq:thermal_state}
\end{align}
We take $N_h>N_c$, or equivalently $\beta_h<\beta_c$.

A type-$k$ collision is generated by the energy-preserving beam-splitter
unitary
\begin{align}
U_k&=\exp\!\left[\theta_k(a^\dagger b_k-a b_k^\dagger)\right],
\qquad 0\leq\theta_k\leq\frac{\pi}{2},
\nonumber\\
\eta_k&=\cos^2\theta_k,\qquad
q_k=1-\eta_k=\sin^2\theta_k.
\label{eq:beam_splitter}
\end{align}

Thus \(\sqrt{\eta_k}=\cos\theta_k\) and \(\sqrt{q_k}=\sin\theta_k\) are nonnegative. The endpoints \(\theta_k=0\) and \(\theta_k=\pi/2\) correspond, respectively, to the identity channel and a complete thermal reset.

Because $[U_k,n+b_k^\dagger b_k]=0$, the energy gained by the system is the
energy lost by the ancilla. Tracing out the outgoing ancilla gives the thermal
attenuator channel
\begin{align}
 \Phi_k(\rho)&=\Tr_{B_k}\!\left[U_k(\rho\otimes\tau_{N_k})U_k^\dagger\right],
 \nonumber\\
 a_{\mathrm{out}}&\equiv U_k^\dagger a U_k
 \nonumber\\
 &=\sqrt{\eta_k}\,a+\sqrt{q_k}\,b_k.
 \label{eq:channel}
\end{align}
The collision duration, which is contained in the integrated angle $\theta_k$,
is conceptually distinct from the arrival rate introduced below.

For $0<\eta_k<1$, the same channel can be embedded in the thermal Lindblad
semigroup
\begin{align}
 \Phi_k&=e^{t_k^{\mathrm{emb}}\mathcal L_k},\qquad
 \eta_k=e^{-\kappa_k t_k^{\mathrm{emb}}},
 \nonumber\\
 \mathcal L_k\rho&=\kappa_k(N_k+1)\D[a]\rho
 +\kappa_kN_k\D[a^\dagger]\rho,
 \label{eq:embedding}
\end{align}
where $t_k^{\mathrm{emb}}$ is the auxiliary semigroup-embedding time, distinct from the physical collision duration, and $\D[L]\rho=L\rho L^\dagger-\{L^\dagger L,\rho\}/2$. Although \(\eta_k=0\) corresponds to the finite collision angle \(\theta_k=\pi/2\), within the semigroup embedding in Eq.~\eqref{eq:embedding} the complete reset is reached only as a limiting channel and not at finite \(t_k^{\mathrm{emb}}\).

\subsection{Poissonian laboratory-time dynamics}

Collisions of type \(k\) occur at the event times of independent homogeneous Poisson processes with intensities \(\lambda_k\). The arrival processes are independent of the system state and of the previously realized collision history. Neglecting the collision duration relative to the mean
waiting time gives the ensemble master equation
\begin{equation}
 \dot\rho=-i\omega[n,\rho]
 +\sum_{k=h,c}\lambda_k(\Phi_k-I)\rho.
 \label{eq:master}
\end{equation}
Each $\Phi_k$ is completely positive and trace preserving (CPTP). In the
interaction picture, the dissipative generator
$\mathcal K=\sum_k\lambda_k(\Phi_k-I)$ is bounded on trace-class operators and
generates the CPTP semigroup
\begin{equation}
 e^{t\mathcal K}=e^{-\Lambda t}\sum_{r=0}^{\infty}
 \frac{t^r}{r!}\left(\sum_k\lambda_k\Phi_k\right)^r,
 \qquad \Lambda=\sum_k\lambda_k.
 \label{eq:poisson_semigroup}
\end{equation}
This is a Poisson mixture of compositions of event channels. The mathematical
form with several event types is known \cite{Nosal2022}; the new results below
come from the thermal-attenuator specialization and its transport counting.

The mean occupation depends only on the effective energy-relaxation rates
\begin{equation}
 \Gamma_k=\lambda_kq_k.
 \label{eq:effective_rate}
\end{equation}
They define the matched continuous reference model
\begin{multline}
 \dot\rho_{\cont}=-i\omega[n,\rho_{\cont}]\\
 +\sum_k\Gamma_k\left[(N_k+1)\D[a]
 +N_k\D[a^\dagger]\right]\rho_{\cont}.
 \label{eq:continuous_reference}
\end{multline}
Gaussian stationary states for an oscillator weakly coupled to a reservoir
with independent thermal components at different temperatures have also
been analyzed microscopically \cite{DodonovMankoManko1995}.

Fig.~\ref{fig:schematic} summarizes the collision model and its matched continuous reference.

\subsection{Observables and heat-counting convention}
We use the normally ordered factorial moments and normalized correlations
\begin{equation}
 F_m=\langle(a^\dagger)^m a^m\rangle,\qquad
 g^{(m)}=\frac{F_m}{\langle n\rangle^m}.
 \label{eq:correlations}
\end{equation}
These quantities can be obtained directly from photon- or phonon-number
statistics.  In the optical setting, photon-number tomography also admits an
explicit reconstruction from measurable homodyne optical tomograms
\cite{MankoManko2009Photon}; probability-representation treatments of a
harmonic oscillator at finite temperature provide related tomographic
descriptions of thermal oscillator states \cite{MankoManko2023Thermal}.
We take the bath-resolved mean heat current to be positive when bath $k$
supplies energy to the system. For heat fluctuations we apply two projective measurements of the
system number immediately before and after every hot collision
\cite{Esposito2009,Campisi2011,BarraLledo2017}. Energy
conservation makes the corresponding increment $\hbar\omega\Delta$, with
$\Delta=m-n$ for a transition $n\to m$, equal to the energy lost by the hot
ancilla. More explicitly, let $\ell$ and $r$ be the measured incoming and
outgoing ancilla occupations. The beam-splitter unitary commutes with
$n+b_h^\dagger b_h$, so every nonzero joint transition probability obeys
\begin{equation}
 m+r=n+\ell,
 \qquad
 \Delta=m-n=\ell-r.
 \label{eq:system_ancilla_tpm_equivalence}
\end{equation}
For an initially number-diagonal system state, every hot and cold collision preserves number diagonality. Let $W_k(m,\Delta|n)$ denote the joint classical kernel of the post-collision system occupation $m$ and the recorded increment $\Delta$ for a type-$k$ event. Equation~\eqref{eq:system_ancilla_tpm_equivalence} shows that resolving the incoming and outgoing ancilla occupations assigns the same increment $\Delta=m-n=\ell-r$ to every allowed joint transition as system-number TPM. After the unrecorded ancilla labels are summed out, both readouts therefore induce the same kernel $W_k$, including the same conditional post-event system population. Iterating these kernels gives the same joint probability law for the complete event-resolved heat record, not only the same single-event marginal. The general operational relation between measuring an energy increment and measuring the initial and final energies is discussed in \cite{Teretenkov2024Superoperator}.

Accordingly, throughout the stationary number-diagonal protocol considered here, the event-resolved heat statistics can be obtained either from system-number TPM at hot events or from measurements of the incoming and outgoing hot-ancilla energies; cold collisions remain unmeasured but preserve number diagonality. The system premeasurement leaves the stationary protocol unchanged because the stationary state below is Fock diagonal. The equivalence also holds throughout a transient that starts from a number-diagonal state. For an initial state with number coherences, however, system TPM defines a measurement-modified transient and need not reproduce the unmeasured coherent evolution. We make no claim about transient heat statistics of such coherent initial states.

\section{Random arrival histories create a thermal mixture}
\label{sec:mixture}

\subsection{Closed hierarchy of normal moments}

The beam-splitter relation in Eq.~\eqref{eq:channel} immediately yields the
following triangular action. A more general single-mode bosonic formula for
diagonal, possibly non-Gaussian environment states is derived in
\cite{Teretenkov2026NonGaussian}; a fermionic counterpart with closed moment
hierarchies is developed in \cite{Teretenkov2026FermionicHierarchies}.

\begin{lemma}[Normal-moment action]
For every integer $m\geq0$,
\begin{equation}
 \Phi_k^*[(a^\dagger)^ma^m]
 =\sum_{j=0}^{m}\binom{m}{j}^2
 \eta_k^{m-j}q_k^j j!N_k^j
 (a^\dagger)^{m-j}a^{m-j}.
 \label{eq:event_moment}
\end{equation}
Consequently, moments of order at most $m$ form a closed system under
Eq.~\eqref{eq:master}.
\end{lemma}

Indeed, thermal averaging removes all terms with unequal numbers of $b_k$ and
$b_k^\dagger$. The surviving ancilla factorial moment is
\begin{equation}
 \langle(b_k^\dagger)^jb_k^j\rangle_{\tau_{N_k}}=j!N_k^j.
 \label{eq:ancilla_factorial_moment}
\end{equation}
Hence
\begin{multline}
 \dot F_m=\sum_k\lambda_k\Bigg[(\eta_k^m-1)F_m\\
 +\sum_{j=1}^{m}\binom{m}{j}^2\eta_k^{m-j}q_k^j
 j!N_k^j F_{m-j}\Bigg].
 \label{eq:moment_ode}
\end{multline}
Eq.~\eqref{eq:moment_ode} is the single-mode thermal specialization of the
general finite hierarchy for Poisson-averaged quadratic dynamics
\cite{Nosal2022}.

\subsection{A random effective occupation}

Thermal attenuators map thermal states to thermal states,
\begin{equation}
 \Phi_k(\tau_x)=\tau_{\eta_kx+q_kN_k}.
 \label{eq:thermal_mapping}
\end{equation}
Conditioned on the sequence of event types, the effective thermal occupation
therefore follows the affine recursion
\begin{equation}
 X_{r+1}=\eta_{K_{r+1}}X_r+q_{K_{r+1}}N_{K_{r+1}},
 \qquad
 \Pr(K=k)=\frac{\lambda_k}{\Lambda}.
 \label{eq:affine_recursion}
\end{equation}
The random variable $X_r$ stores the arrival history through a single scalar.
Because the total waiting rate $\Lambda$ is independent of the system state,
the invariant law of this embedded event chain is also the stationary law in
laboratory time. This is a contractive iterated random function of the standard
perpetuity type \cite{Vervaat1979,Diaconis1999}.

Let
\begin{equation*}
N_{\min}=\min_{k:\lambda_k>0} N_k,
\qquad
N_{\max}=\max_{k:\lambda_k>0} N_k.
\end{equation*}

\begin{theorem}[Stationary thermal mixture]
\label{thm:mixture}
Assume $\lambda_k>0$ and $0\leq\eta_k<1$ for each active event type. The random affine recursion in Eq.~\eqref{eq:affine_recursion} has a unique
invariant probability measure $P_{\ssstate}$ with compact support,
\begin{equation*}
\operatorname{supp} P_{\ssstate}
\subseteq [N_{\min},N_{\max}],
\end{equation*}
satisfying
\begin{equation}
 X\overset{d}{=}\eta_KX'+q_KN_K,
 \label{eq:perpetuity}
\end{equation}
where $X'$ is an independent copy of $X$. The quantum stationary state is
\begin{equation}
\rho_{\ssstate}
=
\int_{N_{\min}}^{N_{\max}}
P_{\ssstate}(dx)\,\tau_x .
\label{eq:stationary_mixture}
\end{equation}
This is the unique stationary density operator of the quantum dynamical
semigroup. For every initial density operator, the symmetrically ordered Weyl
characteristic function converges pointwise to that of
Eq.~\eqref{eq:stationary_mixture}. No trace-norm convergence or convergence rate is asserted.
\end{theorem}

The proof is given in Appendix~\ref{app:mixture}. The essential contraction is
$|X_r-X'_r|=|X_0-X'_0|\prod_{s=1}^r\eta_{K_s}\to0$. The same product is the
total channel transmissivity, so the initial quantum characteristic function
is evaluated at an argument tending to zero. The free rotation commutes with
every phase-covariant thermal attenuator and only rotates that argument by
$e^{-i\omega t}$; it does not change its vanishing modulus. Hence initial
number coherences are erased in the stated pointwise characteristic-function
sense. The stationary mixture is number diagonal, so the Hamiltonian term
vanishes on it. Finally, characteristic functions determine density operators
uniquely, which gives uniqueness within the class of density operators. These
statements deliberately stop short of trace-norm convergence.

The characteristic-function and moment statements use different convergence arguments. Pointwise convergence of Weyl characteristic functions identifies the limiting density operator but does not by itself imply convergence of unbounded number moments. For every fixed order, the normally ordered moments used below converge independently through the finite triangular system in Eq.~\eqref{eq:moment_ode}, provided the corresponding initial moments are finite. Trace-norm convergence, an energy-constrained bound, and a uniform convergence rate are not established.

Eq.~\eqref{eq:stationary_mixture} is generally not a single Gaussian state, but it is a convex mixture of thermal Gaussian states. Its non-Gaussianity therefore arises from convex mixing within this thermal-state representation; no claim is made about its status in a quantum-resource classification.

\subsection{All stationary correlations from the perpetuity}

Let $R_m=\E[X^m]$ and $R_0=1$. Expanding Eq.~\eqref{eq:perpetuity} gives the
following all-order solution.

\begin{theorem}[Stationary moment recursion]
\label{thm:moments}
For every integer $m\geq1$,
\begin{equation}
 R_m=\frac{\displaystyle
 \sum_k\lambda_k\sum_{r=0}^{m-1}\binom{m}{r}
 \eta_k^r(q_kN_k)^{m-r}R_r}
 {\displaystyle\sum_k\lambda_k(1-\eta_k^m)}.
 \label{eq:raw_recursion}
\end{equation}
The stationary quantum factorial moments and normalized correlations are
\begin{equation}
 F_m^{\ssstate}=m!R_m,
 \qquad
 g^{(m)}=\frac{m!R_m}{R_1^m}.
 \label{eq:factorial_from_R}
\end{equation}
\end{theorem}

This representation is more than a computational shortcut: it identifies
fluctuations of the conditional thermal occupation $X$ as the mechanism by
which the arrival history survives stationary averaging.

\section{Average heat flow is blind to individual collisions}
\label{sec:mean_transport}

Let $\mu(t)=\langle n\rangle_t$. The mean-occupation equation obtained from
Eq.~\eqref{eq:moment_ode} is
\begin{equation}
 \dot\mu(t)=-A_1\mu(t)+\sum_k\Gamma_kN_k,
 \qquad
 A_1=\sum_k\Gamma_k.
 \label{eq:mean_dynamics}
\end{equation}
Its stationary value, denoted henceforth by $\mu$, is
\begin{equation}
 \mu\equiv\langle n\rangle_{\ssstate}
 =\frac{\sum_k\Gamma_kN_k}{\sum_k\Gamma_k}.
 \label{eq:mean_ss}
\end{equation}
For a collision of type $k$, the conditional mean number increment is
\begin{equation}
 \E[\Delta|n,k]=q_k(N_k-n).
 \label{eq:conditional_increment}
\end{equation}
Averaging Eq.~\eqref{eq:conditional_increment} over the system state gives
the bath-resolved mean heat current at time $t$,
\begin{equation}
 J_k(t)=\hbar\omega\Gamma_k\left[N_k-\mu(t)\right].
 \label{eq:current_general}
\end{equation}
At stationarity,
\begin{equation*}
 J_k^{\ssstate}=\hbar\omega\Gamma_k(N_k-\mu).
\end{equation*}
For two baths,
\begin{equation}
 J_h^{\ssstate}=-J_c^{\ssstate}=\hbar\omega
 \frac{\Gamma_h\Gamma_c}{\Gamma_h+\Gamma_c}(N_h-N_c).
 \label{eq:two_bath_current}
\end{equation}

Eqs.~\eqref{eq:mean_dynamics} and \eqref{eq:current_general} coincide
with those of the matched continuous model in Eq.~\eqref{eq:continuous_reference}
for the same initial mean occupation. In particular,
Eqs.~\eqref{eq:mean_ss} and \eqref{eq:two_bath_current} give identical
stationary mean occupation and heat currents. Thus neither
the stationary energy nor the mean heat current can determine the individual
$q_k$ and $\lambda_k$ when the products $\Gamma_k$ are held fixed. This is the
null result against which the higher-order signatures below should be read.

Thermodynamically consistent heat-current calculations have been developed
using perturbative corrections to local master equations for weakly coupled
quantum networks \cite{TrushechkinVolovich2016} and the effective Hamiltonian
method for two resonantly interacting oscillators coupled to separate heat
baths \cite{TrubilkoBasharov2019}.
For the present energy-preserving collision model, the direction of average
heat flow is thermodynamically consistent. At steady state,
\begin{equation}
\dot\Sigma=(\beta_c-\beta_h)J_h^{\ssstate}\geq0.
 \label{eq:mean_entropy_production}
\end{equation}
With the convention above, $J_h^{\ssstate}>0$ for $N_h>N_c$, which is equivalent to
$\beta_h<\beta_c$ at the common mode frequency; the sign in
Eq.~\eqref{eq:mean_entropy_production} is therefore positive. This statement
concerns the average entropy production only. It is not a trajectory-level
fluctuation theorem. For broader treatments of entropy production in repeated interactions and
quantum maps, see \cite{Strasberg2017,BarraLledo2017,
LandiPaternostro2021}.

\section{Superthermal correlations retain collision-strength information}
\label{sec:correlations}

\begin{figure*}[t]
 \includegraphics[width=0.98\textwidth]{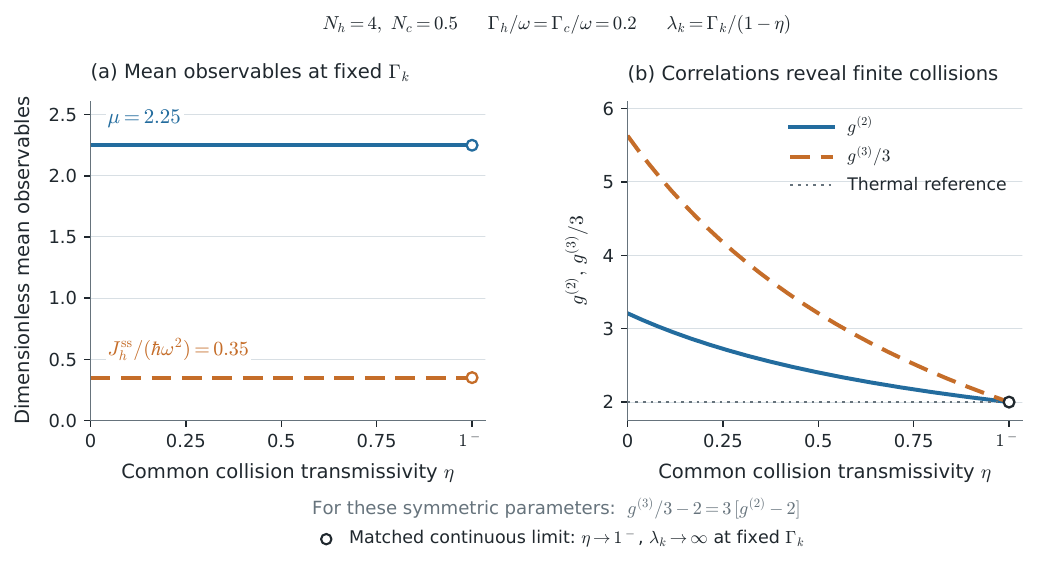}
\caption{Mean observables and stationary correlations at fixed effective rates. The reservoirs have mean occupations \(N_h=4\), \(N_c=0.5\), effective rates \(\Gamma_h/\omega=\Gamma_c/\omega=0.2\), and common collision transmissivity \(\eta_h=\eta_c=\eta\). As \(\eta\) varies, the Poisson arrival rates satisfy \(\lambda_k=\Gamma_k/(1-\eta)\). (a) The stationary occupation \(\mu=2.25\) and dimensionless hot heat current \(J_h^{\mathrm{ss}}/(\hbar\omega^2)=0.35\) remain unchanged. (b) The stationary correlations \(g^{(2)}\) (solid) and \(g^{(3)}/3\) (dashed) approach the common thermal reference value 2 (dotted). Dividing \(g^{(3)}\) by 3 places its thermal value 6 at the same ordinate as \(g^{(2)}=2\). For these symmetric parameters, \(g^{(3)}/3-2=3[g^{(2)}-2]\). Colored curves follow from the exact stationary moment equations. Open circles mark the analytical matched continuous limit \(\eta\to1^-\), \(\lambda_k\to\infty\), at fixed \(\Gamma_k\).}
 \label{fig:transport_correlations}
\end{figure*}

Fig.~\ref{fig:transport_correlations} contrasts the unchanged mean observables
with the varying stationary correlations as $\eta$ changes at fixed effective rates.

Write $Y=X-\mu$ and $c_m=\E[Y^m]$. The centered version of
Eq.~\eqref{eq:raw_recursion} is
\begin{equation}
 c_m=\frac{\displaystyle
 \sum_k\lambda_k\sum_{r=0}^{m-1}\binom{m}{r}\eta_k^r
 [q_k(N_k-\mu)]^{m-r}c_r}
 {\displaystyle\sum_k\lambda_k(1-\eta_k^m)}.
 \label{eq:centered_recursion}
\end{equation}
Here $c_0=1$ and $c_1=0$.
In particular,
\begin{equation}
 v\equiv\Var(X)=
 \frac{\sum_k\lambda_kq_k^2(N_k-\mu)^2}
 {\sum_k\lambda_k(1-\eta_k^2)}.
 \label{eq:v_general}
\end{equation}

\begin{theorem}[Transport-equivalent superthermal separation]
\label{thm:superthermal}
For the finite-collision stationary state,
\begin{equation}
 \Var(n)=\mu(\mu+1)+2v,
 \qquad
 g^{(2)}=2\left(1+\frac{v}{\mu^2}\right).
 \label{eq:g2_general}
\end{equation}
Assume $\lambda_k>0$ and $q_k>0$ for every active event type. Then $v=0$ if and only if $N_k=\mu$ for every active $k$, equivalently, if and only if all active bath occupations coincide. Otherwise $v>0$ and $g^{(2)}>2$, although $\mu$ and $J_h^{\ssstate}$ equal those of the matched continuous model.
\end{theorem}

The result follows from $F_2=2\E[X^2]$ and
$n^2=(a^\dagger)^2a^2+n$. Positivity and the equality condition follow directly
from Eq.~\eqref{eq:v_general}. The law of total variance makes the mechanism
explicit:
\begin{equation}
 \Var(n)=
 \underbrace{\mu^2+\mu+v}_{\E[\Var(n|X)]}
 +\underbrace{v}_{\Var(\E[n|X])}.
 \label{eq:variance_decomposition}
\end{equation}
The arrival history contributes once through fluctuations of the conditional
mean and once through the $X$ dependence of the conditional thermal variance.

For two baths with a common transmissivity $\eta_h=\eta_c=\eta$, define
\begin{equation}
 p=\frac{\Gamma_h}{\Gamma_h+\Gamma_c},
 \qquad
 \mu=pN_h+(1-p)N_c.
 \label{eq:p_definition}
\end{equation}
Then Eq.~\eqref{eq:v_general} reduces to
\begin{equation}
 v=\frac{1-\eta}{1+\eta}
 p(1-p)(N_h-N_c)^2,
 \label{eq:v_common_eta}
\end{equation}
and the central bunching law is
\begin{equation}
 g^{(2)}=2\left[1+\frac{1-\eta}{1+\eta}
 p(1-p)\frac{(N_h-N_c)^2}{\mu^2}\right].
 \label{eq:g2_common_eta}
\end{equation}
At fixed $\Gamma_k$, Eq.~\eqref{eq:g2_common_eta} interpolates monotonically
from a last-collision thermal mixture at $\eta=0$ to the thermal value $2$ as
$\eta\to1$. The higher $g^{(m)}$ follow from
Eq.~\eqref{eq:raw_recursion} and approach $m!$ in the same weak-collision limit.

The excess \(g^{(2)}-2\) is not a signature of nonclassical light. It is generated entirely by classical fluctuations of the conditional thermal occupation \(X\): every state conditioned on a collision history is thermal, and the excess bunching is proportional to \(\Var(X)\).

\subsection{What can and cannot be identified from correlations}
Assume that the bath occupations \(N_h,N_c\) and the effective rates \(\Gamma_h,\Gamma_c\) have been calibrated from mean-occupation relaxation and mean-energy data. Within the common-transmissivity family \(\eta_h=\eta_c=\eta\), these mean observables determine \(p=\Gamma_h/(\Gamma_h+\Gamma_c)\) and \(\mu\), but contain no further information about \(\eta\). By contrast, Eq.~\eqref{eq:g2_common_eta} shows that, for \(N_h\neq N_c\) and \(0<p<1\), the excess bunching \(g^{(2)}-2\) is a strictly decreasing function of the common transmissivity. To make this dependence explicit, define

\begin{equation}
 \mathcal A=p(1-p)\frac{(N_h-N_c)^2}{\mu^2},
 \qquad
 y=\frac{g^{(2)}/2-1}{\mathcal A}.
 \label{eq:identifiability_variables}
\end{equation}
If $p$, $N_h$, and $N_c$ are known and $\mathcal A>0$, then
Eq.~\eqref{eq:g2_common_eta} can be inverted exactly:
\begin{equation}
 \widehat\eta=\frac{1-\widehat y}{1+\widehat y},
 \qquad 0\leq y\leq1.
 \label{eq:eta_estimator}
\end{equation}
Thus, within this calibrated one-parameter family, stationary photon- or phonon-number correlations identify the common transmissivity even though the mean occupation and mean heat current are unchanged.
For a small standard error $\sigma_{g^{(2)}}$ and fixed calibration parameters,
first-order error propagation gives
\begin{equation}
 \sigma_\eta\simeq
 \frac{(1+\eta)^2}{4\mathcal A}\,\sigma_{g^{(2)}}.
 \label{eq:eta_error_propagation}
\end{equation}
This exposes both the operational content and the limitation of the result:
estimation becomes ill conditioned when the bath contrast or one stream weight
is small, and calibration uncertainty in $p$ and $N_k$ must be added in an
experiment-specific analysis.

One value of $g^{(2)}$ does not identify two independent transmissivities. At
fixed $\Gamma_k$, write $\delta_k=N_k-\mu$. Eq.~\eqref{eq:v_general}
becomes
\begin{equation}
 v=\frac{\sum_k\Gamma_kq_k\delta_k^2}
 {\sum_k\Gamma_k(1+\eta_k)},
 \label{eq:v_asymmetric_identifiability}
\end{equation}
so a measured $g^{(2)}$ specifies only a level curve in the
$(\eta_h,\eta_c)$ plane. A second independent stationary statistic can remove
this ambiguity at generic operating points. For example, the third centered
moment is
\begin{equation}
 c_3=\frac{\sum_k\Gamma_k
 [q_k^2\delta_k^3+3\eta_k^2\delta_kv]}
 {\sum_k\Gamma_k(1+\eta_k+\eta_k^2)},
 \label{eq:c3_asymmetric_identifiability}
\end{equation}
and $g^{(3)}=6(\mu^3+3\mu v+c_3)/\mu^3$. Thus the pair
$(g^{(2)},g^{(3)})$, optionally supplemented by $S_h$, can be tested for local
identifiability through its parameter Jacobian. We do not claim that this
Jacobian is nonzero for every parameter choice.

Table~\ref{tab:distinguish} summarizes which observables separate the two
reservoir descriptions.

\begin{table}[b]
 \caption{Observables of the finite Poisson bath and matched continuous bath.
 $R_m$ is defined by Eq.~\eqref{eq:raw_recursion}.}
 \label{tab:distinguish}
 \begin{ruledtabular}
 \begin{tabular}{lcc}
 Observable & finite collisions & continuous bath \\
 \hline
 $\mu$ & Eq.~\eqref{eq:mean_ss} & same \\
 $J_h^{\ssstate}$ & Eq.~\eqref{eq:two_bath_current} & same \\
 $g^{(2)}$ & $2(1+v/\mu^2)$ & $2$ \\
 $g^{(m)}$ & $m!R_m/\mu^m$ & $m!$ \\
 $S_h$ (TPM record) & Eq.~\eqref{eq:heat_noise} & Eq.~\eqref{eq:continuous_noise} \\
 \end{tabular}
 \end{ruledtabular}
\end{table}

\section{Heat noise separates local events from temporal correlations}
\label{sec:heat_noise}

Throughout this section, \(S_h\) refers to an ideal stationary event-resolved TPM record: the system starts in its Fock-diagonal stationary state, every hot event is identified, and the relevant system or ancilla occupations are measured with ideal number resolution. Missed collisions, finite detector bandwidth, detection inefficiency, and state-preparation errors are not included. In this number-diagonal setting the same record defines a natural classical jump-trajectory unraveling. We do not claim unraveling-independent heat statistics for coherent initial states or for an unmeasured quantum process.

\subsection{Counting deformation and cumulant rates}

Let $P_k(m|n)$ be the population transition probability of the type-$k$
thermal attenuator, and let $Q_h(t)$ be the cumulative hot-stream increment in
boson-number units. A convenient finite-cutoff bookkeeping deformation is
\begin{align}
 K_\chi&=\lambda_h(P_h^{(\chi)}-I)+\lambda_c(P_c-I),
 \nonumber\\
 P_h^{(\chi)}(m|n)&=e^{\chi(m-n)}P_h(m|n).
 \label{eq:tilted_generator}
\end{align}
This is standard in counting statistics \cite{Esposito2009}. In the analytical
derivation below we do not assume that $K_\chi$ has an isolated dominant
eigenvalue on an infinite-dimensional population space. Instead, we define
only the two stationary long-time rates
\begin{align}
 j_h&=\lim_{t\to\infty}\frac{\E Q_h(t)}{t},\qquad
 s_h=\lim_{t\to\infty}\frac{\Var Q_h(t)}{t},
 \nonumber\\
J_h^{\ssstate}&=\hbar\omega j_h,\qquad
S_h=(\hbar\omega)^2s_h.
 \label{eq:cumulants}
\end{align}
They follow from closed equations for joint moments of $Q_h(t)$ and $n(t)$.
No finite-$\chi$ spectral statement is made.

For an incoming Fock state $|n\rangle$, the first two one-event moments of
$\Delta=m-n$ are
\begin{equation}
 g_h(n)\equiv\E[\Delta|n,h]=q_h(N_h-n),
 \label{eq:event_heat_mean}
\end{equation}
and
\begin{multline}
 h_h(n)\equiv\E[\Delta^2|n,h]
 =q_h^2\left[(N_h-n)^2+N_h(N_h+1)\right]\\
 +\eta_hq_h(2nN_h+n+N_h).
 \label{eq:event_heat_second}
\end{multline}
These one-event moments are derived in Appendix~\ref{app:fcs}.
The last line contains bosonic partition noise, while the first line contains
the occupation mismatch and thermal number noise.

\subsection{Exact zero-frequency noise}

\begin{figure*}[t]
 \includegraphics[width=\textwidth]{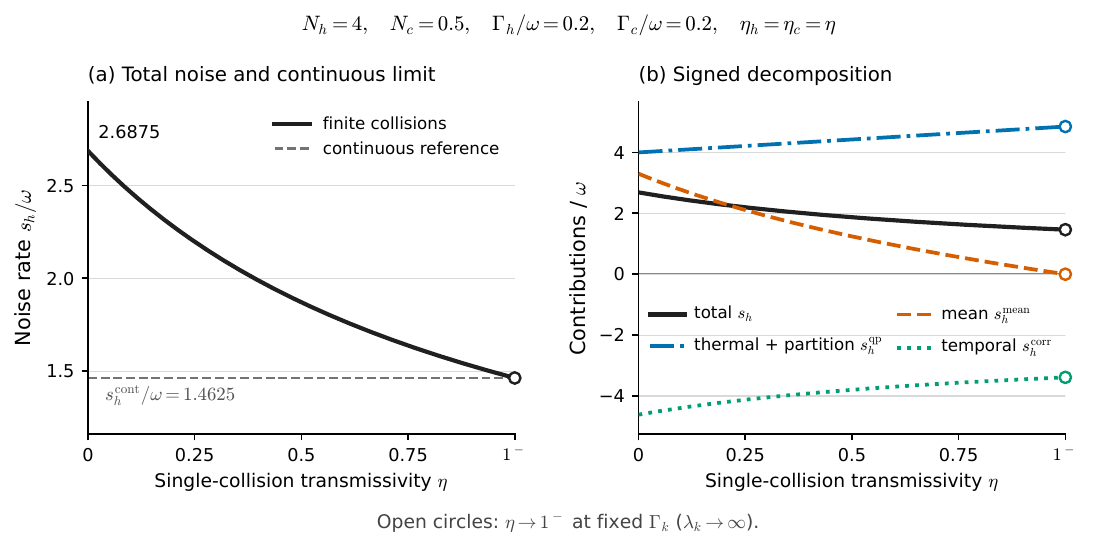}
 \caption{Hot-stream heat noise of the ideal stationary event-resolved TPM record for \(N_h=4\), \(N_c=0.5\), and \(\eta_h=\eta_c=\eta\), at fixed effective rates \(\Gamma_k=\lambda_k(1-\eta)\), with \(\Gamma_h/\omega=\Gamma_c/\omega=0.2\). All noise rates are plotted in units of \(\omega\), where \(S_h=(\hbar\omega)^2s_h\). (a) Exact total noise (solid black) and the matched continuous-bath reference (dashed gray), \(s_h^{\mathrm{cont}}/\omega=1.4625\). (b) Additive decomposition \(s_h=s_h^{\mathrm{loc,mean}}+s_h^{\mathrm{loc,var}}+s_h^{\mathrm{cov}}\) into the local conditional-mean term (dashed orange), the local conditional-variance term containing thermal and beam-splitter partition fluctuations (dash-dotted blue), and the covariance-response term (dotted green). The solid black curve shows their sum. The first two contributions are nonnegative, whereas the covariance-response term is negative for this parameter family. These algebraic contributions are not independent noise spectra. Open circles indicate analytic limits as \(\eta\to1^-\) and \(\lambda_k\to\infty\) at fixed \(\Gamma_k\); \(s_h^{\mathrm{cov}}/\omega\to-3.3875\).}
 \label{fig:heat_noise}
\end{figure*}

Fig.~\ref{fig:heat_noise} shows the exact heat noise and its decomposition
into local and temporal contributions.

Define
\begin{align}
 M_2&=\langle n^2\rangle_{\ssstate}=2(\mu^2+v)+\mu,
 \nonumber\\
 \overline h_h&=\langle h_h(n)\rangle_{\ssstate},\qquad
 \langle ng_h\rangle=q_h(N_h\mu-M_2),
 \label{eq:heat_auxiliary}
\end{align}
and
\begin{equation}
 C_h(t)=\operatorname{Cov}[Q_h(t),n(t)].
 \label{eq:heat_covariance}
\end{equation}
When the population process starts in its stationary state, direct conditioning
on the next hot or cold event gives
\begin{equation}
 \dot C_h(t)=-A_1C_h(t)
 +\lambda_h(\langle ng_h\rangle+\overline h_h)-j_h\mu.
 \label{eq:heat_covariance_ode}
\end{equation}
Consequently,
\begin{equation}
 z_h\equiv\lim_{t\to\infty}C_h(t)
 =\frac{\lambda_h(\langle ng_h\rangle+\overline h_h)-j_h\mu}{A_1}.
 \label{eq:z_definition}
\end{equation}

\begin{figure*}[t]
 \includegraphics[width=\textwidth]{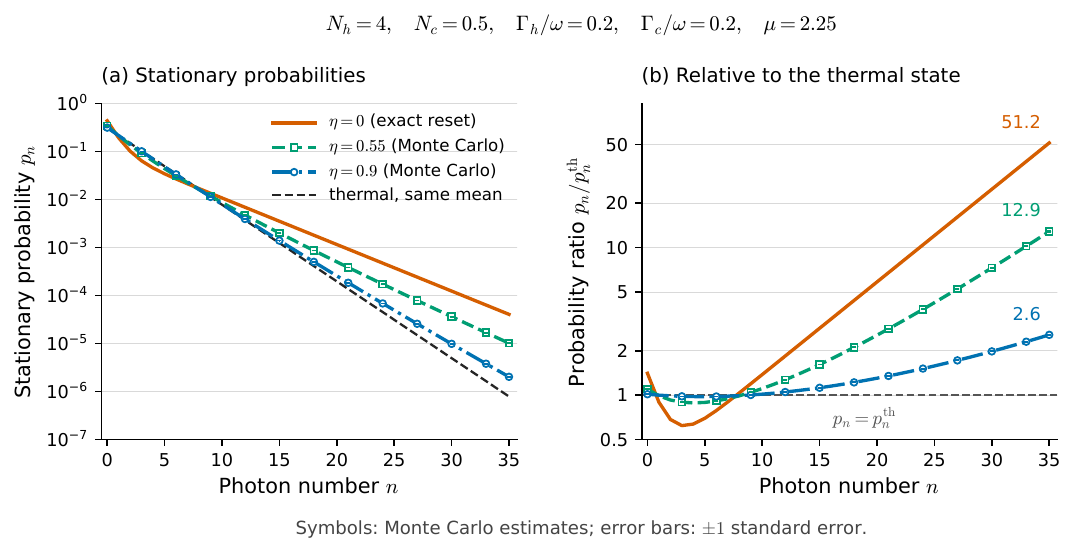}
 \caption{Stationary photon-number distribution of the collision model for $N_h=4$, $N_c=0.5$, and $\eta_h=\eta_c=\eta$, at fixed effective rates $\Gamma_k=\lambda_k(1-\eta)$, with $\Gamma_h/\omega=0.2$ and $\Gamma_c/\omega=0.2$. All cases have the same exact mean occupation $\mu=2.25$. (a) Probabilities $p_n$ and the matched thermal reference $p_n^{\mathrm{th}}=\mu^n/(1+\mu)^{n+1}$ (dashed black). (b) Ratios $p_n/p_n^{\mathrm{th}}$; the dashed horizontal line denotes the thermal reference. Both vertical axes are logarithmic. The $\eta=0$ curve is evaluated analytically from the full-reset mixture $p_n=\sum_k(\lambda_k/\Lambda)N_k^n/(1+N_k)^{n+1}$, where $\Lambda=\lambda_h+\lambda_c$. For $\eta=0.55$ and $0.9$, probabilities are Monte Carlo averages of $X^n/(1+X)^{n+1}$ over $10^5$ independent samples, each obtained after $220$ burn-in iterations of the affine recursion. Symbols and error bars, displayed at selected integer $n$, show these estimates and one standard error; the bars are smaller than many symbols. Lines connect integer-$n$ values as visual guides. For the displayed parameters, stronger collisions enhance the large-$n$ probabilities while reducing probabilities at intermediate $n$.}
 \label{fig:photon_distribution}
\end{figure*}

\begin{theorem}[Asymptotic rates of the first two heat cumulants]
\label{thm:heat_noise}
For the ideal stationary, number-diagonal, event-resolved TPM convention of
Eq.~\eqref{eq:system_ancilla_tpm_equivalence},
\begin{equation}
 j_h=\lambda_hq_h(N_h-\mu),
 \qquad
 s_h=\lambda_h\overline h_h-2\lambda_hq_hz_h.
 \label{eq:heat_noise}
\end{equation}
Thus the exact zero-frequency heat noise depends only on the first two
stationary occupation moments and the scalar response moment $z_h$.
\end{theorem}

\begin{figure*}[t]
 \includegraphics[width=\textwidth]{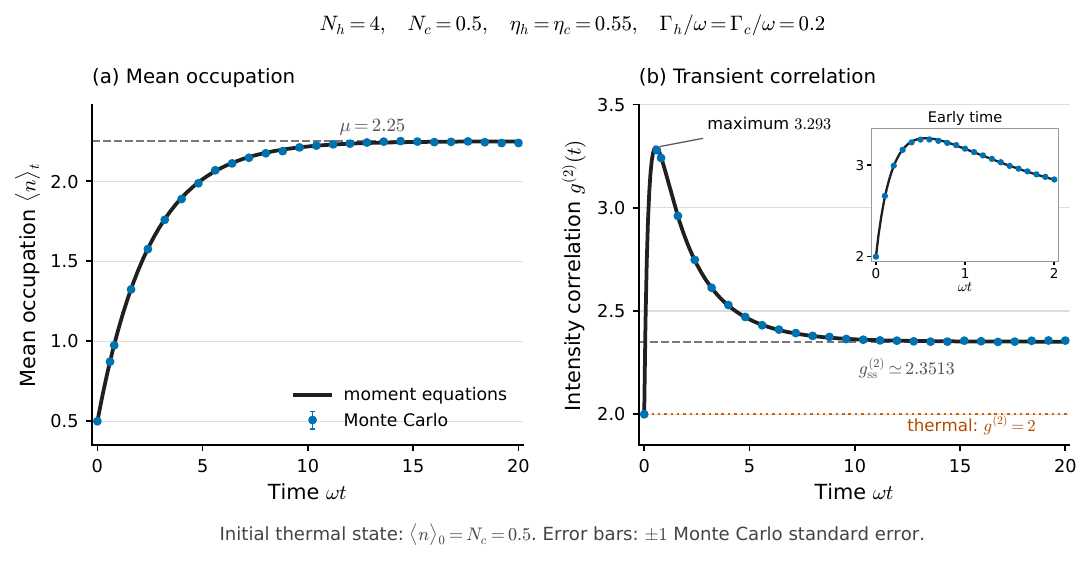}
\caption{Transient mean occupation and intensity correlation for $N_h=4$, $N_c=0.5$, $\eta_h=\eta_c=0.55$, and $\Gamma_h/\omega=\Gamma_c/\omega=0.2$, starting from the thermal state $\tau_{N_c}$. (a) Mean occupation $\langle n\rangle_t$, approaching $\mu=2.25$. (b) Correlation $g^{(2)}(t)$, approaching the nonthermal stationary value $g^{(2)}_{\mathrm{ss}}\simeq2.3513$. Solid black curves solve the closed moment equations; blue circles are Monte Carlo estimates from $3\times10^4$ independent realizations of the affine process $X_t$. Gray dashed lines mark the stationary values, and the orange dotted line in (b) marks the thermal value $g^{(2)}=2$. The inset resolves $0\leq\omega t\leq2$. For these parameters the transient maximum is $g^{(2)}\simeq3.2928$ at $\omega t\simeq0.5568$. Error bars give pointwise Monte Carlo standard errors across independent trajectories, using the delta method for $g^{(2)}$; some are smaller than the plotting symbols.}
 \label{fig:transient}
\end{figure*}

Indeed, the exact variance balance is
\begin{equation}
 \frac{d}{dt}\Var Q_h(t)
 =\lambda_h\overline h_h-2\lambda_hq_hC_h(t),
 \label{eq:heat_variance_balance}
\end{equation}
which proves the theorem after taking the long-time limit.
Appendix~\ref{app:fcs} gives the joint-moment derivation in detail. All terms
are finite because the stationary thermal mixture has moments of every order
for finite $N_k$. The result can be decomposed as
\begin{equation}
 s_h=s_h^{\mathrm{loc,mean}}+s_h^{\mathrm{loc,var}}+s_h^{\mathrm{cov}},
 \label{eq:noise_decomposition}
\end{equation}
where
\begin{align}
 s_h^{\mathrm{loc,mean}}
 &=\lambda_hq_h^2\left[(N_h-\mu)^2+\Var(n)\right],
 \label{eq:noise_mean_part}\\
 s_h^{\mathrm{loc,var}}
 &=\lambda_h\left[q_h^2N_h(N_h+1)
 +\eta_hq_h(2N_h\mu+\mu+N_h)\right],
 \label{eq:noise_qp_part}\\
 s_h^{\mathrm{cov}}&=-2\lambda_hq_hz_h.
 \label{eq:noise_corr_part}
\end{align}
The first two terms are nonnegative local one-event contributions. The term \(s_h^{\mathrm{loc,mean}}\) is generated by the conditional mean increment, whereas \(s_h^{\mathrm{loc,var}}\) contains the thermal and beam-splitter partition contributions to the conditional event variance. The covariance-response term \(s_h^{\mathrm{cov}}\) contains the effect of correlations between the accumulated heat record and the mode occupation. In additive-functional language, the conditional event variance contributes to the predictable quadratic variation of the martingale part of the heat record, whereas \(s_h^{\mathrm{cov}}\) contains the integrated temporal response mediated by the system occupation. This gives a Green--Kubo-like interpretation, but not a unique decomposition into experimentally separable noise sources. These separate terms are algebraic contributions rather than independently measurable noise spectra; only their sum \(s_h\), which is nonnegative by definition as a variance-growth rate, is the physical zero-frequency noise. We do not assert that \(s_h^{\mathrm{cov}}\) is negative outside the parameter family displayed in Fig.~\ref{fig:heat_noise}.

Operationally, \(S_h\) is the variance-growth rate of the specified hot-stream heat record. By Eq.~\eqref{eq:system_ancilla_tpm_equivalence}, the incoming/outgoing ancilla record is equivalent to the system-number TPM record under the stationary number-diagonal protocol considered here.

\subsection{Strong- and weak-collision limits}

For a full reset, $\eta_k=0$, the type of the last event completely determines
the conditional thermal state. For two baths,
\begin{equation}
 P_{\ssstate}(dx)=\frac{\lambda_h}{\Lambda}\delta(x-N_h)
 +\frac{\lambda_c}{\Lambda}\delta(x-N_c).
 \label{eq:reset_law}
\end{equation}
At the opposite extreme, take $q_k\to0$ and $\lambda_k\to\infty$ at fixed $\Gamma_k=\lambda_kq_k$. For every fixed moment order $m$, the adjoint generator converges coefficientwise on the finite-dimensional space $\operatorname{span}\{(a^\dagger)^r a^r:0\leq r\leq m\}$ to the corresponding thermal Lindblad moment generator. In this restricted sense,
\begin{equation}
 \lambda_k(\Phi_k^*-I)O\longrightarrow
 \Gamma_k[(N_k+1)\D^*[a]+N_k\D^*[a^\dagger]]O.
 \label{eq:weak_generator}
\end{equation}
The channel expansion underlying this limit is derived in
Appendix~\ref{app:weak_limit}.
$v=O(q)$, $g^{(m)}\to m!$, and Eq.~\eqref{eq:heat_noise} reduces to the
continuous-cavity result. By contrast, multiplying every $\lambda_k$ by the
same constant at fixed $\eta_k$ leaves the embedded probabilities
$\lambda_k/\Lambda$ and hence the invariant affine law unchanged; it multiplies
the generator by that constant and changes only the laboratory-time scale. It
therefore does not Gaussianize the stationary state. The point $\eta=1$ at
nonzero fixed $\Gamma_k$ is never a finite-rate Poisson model: it denotes the
joint limit $q_k\to0$, $\lambda_k\to\infty$.

\section{Collision-strength crossover in observable statistics}
\label{sec:numerics}

We evaluate the exact recursions for $N_h=4$, $N_c=0.5$, and
$\Gamma_h/\omega=\Gamma_c/\omega=0.2$, keeping $\Gamma_k/\omega$ fixed while changing the common
$\eta$. We set $\hbar=\omega=1$ in all numerical data. Selected values are
listed in Table~\ref{tab:selected_values}. The stationary mean occupation and
heat current are independent of the common transmissivity $\eta$, whereas all
displayed higher correlations and the heat noise change strongly.

Fig.~\ref{fig:photon_distribution} shows the corresponding stationary photon-number distributions.

\begin{table*}[t]
 \caption{Stationary observables at fixed effective rates. Thermal values at
 the same mean are $g^{(2)}=2$, $g^{(3)}=6$, and $g^{(4)}=24$.}
 \label{tab:selected_values}
 \begin{ruledtabular}
 \begin{tabular}{cccccc}
 $\eta$ & $g^{(2)}$ & $g^{(3)}$ & $g^{(4)}$
& $J_h^{\ssstate}/(\hbar\omega^2)$ & $S_h/[(\hbar\omega)^2\omega]$ \\
 \hline
 0.00 & 3.2098765 & 16.8888889 & 119.8939186 & 0.3500000 & 2.6875000 \\
 0.55 & 2.3512545 &  9.1612903 &  50.7183066 & 0.3500000 & 1.8181452 \\
 0.90 & 2.0636777 &  6.5730994 &  28.6526749 & 0.3500000 & 1.5269737 \\
 0.98 & 2.0122210 &  6.1099888 &  24.8825624 & 0.3500000 & 1.4748737 \\
 \end{tabular}
 \end{ruledtabular}
\end{table*}

Photon-number distributions for general one-mode Gaussian states, including
thermal states, were obtained in Ref.~\cite{DodonovMankoManko1994}.
In the present model, the stationary state is generally non-Gaussian.
Averaging the geometric distribution of each thermal component in
Eq.~\eqref{eq:stationary_mixture} gives
\begin{equation}
 p_n=\int P_{\ssstate}(dx)\frac{x^n}{(1+x)^{n+1}}.
 \label{eq:photon_distribution}
\end{equation}
Fig.~\ref{fig:photon_distribution} shows a broader distribution for stronger
collisions than for the thermal state with the same mean. The displayed tails
are numerical results for the stated parameter set; we do not infer a universal
tail theorem from them.

\subsection{Independent trajectory and Fock-space checks}

The affine process in Eq.~\eqref{eq:affine_recursion} is sampled directly with
exact Poisson event counts, not an Euler approximation. Fig.~\ref{fig:transient}
compares $3\times10^4$ trajectories with the exact moment ODE for a cold
initial thermal state and $\eta=0.55$. The observed transient overshoot of
$g^{(2)}(t)$ is a feature of this parameter set, not a theorem about all cold
initial states. At early times the random hot/cold histories broaden the distribution of \(X_t\) faster than its mean approaches stationarity; subsequent contractions suppress this relative dispersion. The resulting overshoot of \(\Var(X_t)/\E[X_t]^2\) is therefore parameter dependent.

We independently construct each population channel $P_k$ in a truncated Fock
basis using a pure-loss channel followed by a quantum-limited amplifier. The
stationary population solves
\begin{equation}
 \left[\sum_k\lambda_k(P_k-I)\right]p_{\ssstate}=0.
 \label{eq:fock_stationary}
\end{equation}
Derivatives at $\chi=0$ of the dominant eigenvalue of the
finite-dimensional tilted generator in Eq.~\eqref{eq:tilted_generator}
give the asymptotic rates of the first two heat cumulants without using
the analytical reduction in Theorem~\ref{thm:heat_noise}; this is a
numerical check, not an infinite-dimensional spectral assumption.
Across $\eta=0,0.55,0.9$ the maximum relative errors in
$(\mu,g^{(2)},J_h^{\ssstate},S_h)$ decrease from
$(1.66\times10^{-5},1.23\times10^{-4},2.14\times10^{-5},1.79\times10^{-4})$
at cutoff 60 to
$(3.07\times10^{-8},3.55\times10^{-7},3.95\times10^{-8},4.86\times10^{-7})$
at cutoff 90 and
$(5.05\times10^{-11},7.95\times10^{-10},6.49\times10^{-11},1.05\times10^{-9})$
at cutoff 120. Appendix~\ref{app:numerics} gives the transition matrices used
in this check.

\section{Conclusion}
\label{sec:conclusion}

We asked whether a finite Poisson collision bath can be distinguished from a
continuous Gaussian reservoir after their mean dynamics have been matched. For
a bosonic mode between hot and cold thermal streams, the answer is yes. The
mean occupation and mean heat current depend only on
$\Gamma_k=\lambda_k(1-\eta_k)$ and are identical in the two descriptions.
Nevertheless, the random sequence of competing thermal fixed points produces a
stationary mixture governed by an affine perpetuity. Its classical fluctuations yield strict superthermal bunching and exact higher-order correlations that remain sensitive to the individual collision strengths; the excess bunching is not a signature of nonclassical light.

The same distinction appears in transferred-energy fluctuations. We derived
the asymptotic rates of the first two heat cumulants and separated the
zero-frequency noise into local occupation-mismatch, thermal/partition, and
temporal-correlation terms. Independent Fock-space diagonalization and Monte Carlo trajectories
validate the stationary formulas and their crossover between full resets and
the continuous weak-collision limit.

The contribution is not the compound-Poisson collision construction itself, but its exact solution in this nonequilibrium bosonic setting. Specifically, we obtain the stationary affine thermal mixture, an all-order recursion for the stationary normally ordered moments, and the finite-event correction to the zero-frequency heat noise. The scope of these results is deliberately restricted. The model describes a single bosonic mode driven by independent Markovian Poisson streams of fresh thermal ancillas through instantaneous, nonoverlapping, energy-preserving collisions. We derive the asymptotic rates of the first two heat cumulants, but not the
full finite-counting-field scaled cumulant generating function or a Gallavotti–Cohen fluctuation symmetry. Global identifiability is not established when the two collision transmissivities vary independently, and a quantitative assessment of experimental precision requires a platform-specific treatment of state preparation, finite sampling, detector bandwidth, and detection efficiency.

The most direct measurable consequence is that stationary number correlations remain sensitive to collision granularity after the bath occupations and effective relaxation rates have been calibrated. Within the common-transmissivity family, the mean occupation and mean heat current contain no additional information about \(\eta\), whereas the excess bunching \(g^{(2)}-2\) identifies the common transmissivity through Eq.~\eqref{eq:eta_estimator}. For independent \(\eta_h\) and \(\eta_c\), the correlations establish sensitivity but not global identifiability. Heat noise provides a complementary but experimentally more demanding probe because it requires event-resolved energy measurements.

Several platforms provide complementary ingredients for a future test, without
singling out a unique implementation.  Programmable beam-splitter interactions
between circuit-QED microwave cavities are available
\cite{Gao2018,Chapman2023}; optical collision simulators and pulsed sources with
thermal photon statistics provide routes based on propagating or time-bin modes
\cite{Cuevas2019,Wakui2025}; and trapped-ion motional modes permit controlled
bosonic thermal reservoirs and occupation measurements \cite{So2026}.  The
construction can also be used modelwise, rather than as a literal sequence of
bosonic beam splitters: refreshed ancillas and a prescribed random gate schedule
define a digital simulation of the collision process.  Experiments on IBM
processors have already implemented collision-model open dynamics and
collision-based collective dissipation \cite{GarciaPerez2020,Cattaneo2023},
although not the specific two-stream bosonic protocol studied here.  A natural
next step is therefore to incorporate finite sampling and realistic detection
into a calibrated implementation. Extending the analysis to the full tilted
spectrum and possible fluctuation symmetries is a separate theoretical direction.

\begin{acknowledgments}
We would like to thank Ranjit Singh for valuable discussions, insightful comments, and helpful feedback that contributed to the development of this work.

This work was supported by the Russian Science Foundation under grant no.~25-21-00700, \url{https://rscf.ru/en/project/25-21-00700/}.
\end{acknowledgments}

\appendix

\section{Channel action and complete positivity}
\label{app:channel}

Eq.~\eqref{eq:channel} is CPTP because it is a unitary dilation followed
by a partial trace. The Poisson expansion in Eq.~\eqref{eq:poisson_semigroup} is a
convex sum of compositions of CPTP maps: writing
$\overline\Phi=\Lambda^{-1}\sum_k\lambda_k\Phi_k$ gives
\begin{equation}
 e^{t\mathcal K}=e^{-\Lambda t}\sum_{r=0}^{\infty}
 \frac{(\Lambda t)^r}{r!}\overline\Phi^r.
 \label{eq:cptp_expansion_appendix}
\end{equation}
The weights are Poisson probabilities and sum to one. Moreover,
$\|\Phi_k-I\|_{1\to1}\leq2$, so $\mathcal K$ is bounded on trace class in the
interaction picture.

To prove Eq.~\eqref{eq:event_moment}, insert
$a_{\mathrm{out}}=\sqrt{\eta_k}a+\sqrt{q_k}b_k$ into
$(a_{\mathrm{out}}^\dagger)^ma_{\mathrm{out}}^m$. The binomial expansion has
terms indexed by $r,s$. Gauge invariance of the thermal ancilla state removes
$r\neq s$, and the surviving terms give
\begin{equation}
 \binom{m}{j}^2\eta_k^{m-j}q_k^j
 \langle(b_k^\dagger)^jb_k^j\rangle
 =\binom{m}{j}^2\eta_k^{m-j}q_k^j j!N_k^j.
\end{equation}

\section{Stationary mixture and moment recursions}
\label{app:mixture}

Take two copies of Eq.~\eqref{eq:affine_recursion} driven by the same event
sequence. Their distance obeys
\begin{equation}
 |X_r-X'_r|=|X_0-X'_0|\prod_{s=1}^{r}\eta_{K_s}.
 \label{eq:coupling_contraction}
\end{equation}
Since every active $\eta_k<1$ and event types are sampled independently with
nonzero probabilities, the product vanishes almost surely. The standard
iterated-random-function coupling therefore gives a unique invariant
probability measure $P_{\ssstate}$. The corresponding invariant random variable
may be represented by the convergent backward series
\begin{equation}
 X\overset{d}{=}\sum_{r=1}^{\infty}
 q_{K_r}N_{K_r}\prod_{s=1}^{r-1}\eta_{K_s}.
 \label{eq:backward_perpetuity}
\end{equation}
Since $q_k=1-\eta_k$, the coefficients in
Eq.~\eqref{eq:backward_perpetuity} are nonnegative and sum to unity almost
surely. Because every $N_k\in[N_{\min},N_{\max}]$, it follows that
\begin{equation*}
 \operatorname{supp}P_{\ssstate}
 \subseteq [N_{\min},N_{\max}].
\end{equation*}

A composition of conditional attenuators has total transmissivity $\eta_{K_r}\cdots\eta_{K_1}$. In the symmetrically ordered characteristic function, the factor carrying the initial system state is evaluated at an argument multiplied by the square root of this transmissivity. Free evolution adds the phase $e^{-i\omega t}$ to that argument but leaves its modulus unchanged. Continuity of the characteristic function at the origin therefore gives $\chi_{\rho_0}(0)=1$ in the event-count limit $r\to\infty$ for every density operator, including states with number coherences. The remaining Gaussian factor is the characteristic function of a thermal state with the random occupation in Eq.~\eqref{eq:backward_perpetuity}. Its modulus is bounded by one, so dominated convergence justifies averaging over event histories and gives Eq.~\eqref{eq:stationary_mixture} together with pointwise characteristic-function convergence.

To pass from event count to laboratory time, let $N_t$ be the total number of events up to time $t$. It is Poisson distributed with mean $\Lambda t$, independently of the event types, so $N_t\to\infty$ in probability and the no-event contribution has weight $e^{-\Lambda t}$. Splitting the Poisson average at an arbitrary fixed event count and then sending first $t\to\infty$ and subsequently the cutoff to infinity proves the same pointwise limit in laboratory time. The state-independent holding rate also implies directly that an invariant law of the embedded chain is invariant for the continuous-time jump generator. Since the limiting mixture commutes with $n$, it is stationary in the presence of the free Hamiltonian. Applying the limiting statement to any stationary density operator and using uniqueness of the Weyl characteristic representation proves quantum-state uniqueness. Establishing a uniform energy-constrained or trace-norm convergence bound is left open.

For the raw moments, raise Eq.~\eqref{eq:perpetuity} to the $m$th power and use
independence of $K$ and $X'$:
\begin{equation}
 R_m=\frac{1}{\Lambda}\sum_k\lambda_k
 \sum_{r=0}^{m}\binom{m}{r}\eta_k^r(q_kN_k)^{m-r}R_r.
 \label{eq:raw_pre_recursion}
\end{equation}
Moving the $r=m$ terms to the left gives
Eq.~\eqref{eq:raw_recursion}. The centered recursion in
Eq.~\eqref{eq:centered_recursion} follows in the same way from
\begin{equation}
 Y\overset{d}{=}\eta_KY'+q_K(N_K-\mu).
 \label{eq:centered_perpetuity}
\end{equation}

\section{Direct joint-moment derivation of heat noise}
\label{app:fcs}

Start the population process in its stationary law and write
$Q\equiv Q_h(t)$. Conditioning on the next event gives
\begin{equation}
 \frac{d}{dt}\E Q=\lambda_h\langle g_h(n)\rangle=j_h.
 \label{eq:direct_mean_heat}
\end{equation}
For the mixed moment, a cold event changes $Qn$ by $Q\Delta$, whereas a hot
event changes it by
$(Q+\Delta)(n+\Delta)-Qn=Q\Delta+n\Delta+\Delta^2$. Hence
\begin{multline}
 \frac{d}{dt}\E[Qn]
 =\sum_k\lambda_k\E[Qg_k(n)]\\
 +\lambda_h\langle ng_h(n)+h_h(n)\rangle.
 \label{eq:direct_joint_moment}
\end{multline}
Because
\begin{equation}
 \sum_k\lambda_kg_k(n)=\sum_k\Gamma_kN_k-A_1n
 =A_1(\mu-n),
 \label{eq:drift_identity}
\end{equation}
subtracting $d(\mu\E Q)/dt=\mu j_h$ from
Eq.~\eqref{eq:direct_joint_moment} gives
Eq.~\eqref{eq:heat_covariance_ode}. Its unique long-time limit is
Eq.~\eqref{eq:z_definition}.

Likewise, a hot event changes $Q^2$ by $2Q\Delta+\Delta^2$, so
\begin{equation}
 \frac{d}{dt}\E Q^2
 =2\lambda_h\E[Qg_h(n)]+\lambda_h\overline h_h.
 \label{eq:direct_second_heat}
\end{equation}
Using $g_h(n)=q_h(N_h-n)$ and subtracting
$d(\E Q)^2/dt$ yields the variance balance in
Eq.~\eqref{eq:heat_variance_balance}. Since $C_h(t)\to z_h$, division by time
proves Eq.~\eqref{eq:heat_noise}. This argument uses only finite stationary
second moments and a scalar stable equation with rate $A_1>0$; it requires no
spectral perturbation of an infinite-dimensional tilted operator.

For completeness, the one-event second moment can be checked directly. The
beam-splitter relation gives
\begin{equation}
 \Delta=q_h(b_h^\dagger b_h-n)
 +\sqrt{\eta_hq_h}(a^\dagger b_h+b_h^\dagger a).
 \label{eq:increment_operator}
\end{equation}
In $|n\rangle\langle n|\otimes\tau_{N_h}$, the cross term has zero mean,
\begin{equation}
 \E[(b_h^\dagger b_h-n)^2]=(N_h-n)^2+N_h(N_h+1),
 \label{eq:thermal_mismatch_second}
\end{equation}
and
\begin{equation}
 \E[(a^\dagger b_h+b_h^\dagger a)^2]
 =n(N_h+1)+(n+1)N_h.
 \label{eq:partition_second}
\end{equation}
Combining these identities gives Eq.~\eqref{eq:event_heat_second} with the
stated sign convention.

\section{Continuous weak-collision limit}
\label{app:weak_limit}

For a fixed $m$, let $\mathcal V_m=\operatorname{span}\{(a^\dagger)^r a^r:0\leq r\leq m\}$. Write $\D^*[L]O=L^\dagger O L-\{L^\dagger L,O\}/2$ for the adjoint dissipator. The exact triangular action in Eq.~\eqref{eq:event_moment} implies, coefficientwise on $\mathcal V_m$,
\begin{equation}
 \Phi_k^*-I=q_k\left[(N_k+1)\D^*[a]+N_k\D^*[a^\dagger]\right]+O_m(q_k^2),
 \label{eq:channel_weak_expansion}
\end{equation}
where the remainder means that every matrix coefficient in this finite-dimensional restriction is $O(q_k^2)$. Thus Eq.~\eqref{eq:channel_weak_expansion} is not asserted as an unrestricted operator-norm expansion on unbounded observables. Taking $\lambda_kq_k=\Gamma_k$ gives Eq.~\eqref{eq:weak_generator} in the corresponding moment sense. For the common-$\eta$ two-bath family,
Eq.~\eqref{eq:v_common_eta} explicitly gives $v=O(q)$, and the triangular
recursion then gives $R_m\to\mu^m$ and $g^{(m)}\to m!$ order by order.

The heat-noise limit may be written
\begin{equation}
 s_h^{\cont}=\Gamma_h[N_h+(2N_h+1)\mu]-2\Gamma_hz_h^{\cont},
 \label{eq:continuous_noise}
\end{equation}
where
\begin{equation}
 z_h^{\cont}=\frac{\Gamma_h}{\Gamma_h+\Gamma_c}
 \left[N_h+2N_h\mu-\mu^2\right].
 \label{eq:continuous_z}
\end{equation}
This agrees with the continuous thermal-cavity counting framework
\cite{Brange2019} under the counting convention used here.

\section{Numerical implementation and reproducibility}
\label{app:numerics}

The analytical curves are generated recursively from
Eqs.~\eqref{eq:raw_recursion}, \eqref{eq:v_general}, and
\eqref{eq:heat_noise}. For an explicit independent population construction,
set
\begin{equation*}
G_k=1+q_kN_k,
\qquad
\eta_k^{\mathrm{loss}}=\frac{\eta_k}{G_k}.
\end{equation*}
The thermal attenuator is a pure-loss channel of transmissivity
$\eta_k^{\mathrm{loss}}$ followed by a quantum-limited amplifier of gain $G_k$,
with matrix elements
\begin{align}
 L_k(\ell|n)
&=\binom{n}{\ell}
(\eta_k^{\mathrm{loss}})^\ell
(1-\eta_k^{\mathrm{loss}})^{n-\ell},
\qquad 0\leq\ell\leq n,
\label{eq:loss_transition}\\
 A_k(m|\ell)
 &=\binom{m}{\ell}G_k^{-(\ell+1)}
 (1-G_k^{-1})^{m-\ell},
 \qquad m\geq\ell,
 \label{eq:amplifier_transition}\\
 P_k(m|n)
 &=\sum_{\ell=0}^{\min(m,n)}A_k(m|\ell)L_k(\ell|n).
 \label{eq:attenuator_transition_explicit}
\end{align}
The stationary vector and finite-matrix counting derivatives are computed at
Fock cutoffs 60, 90, and 120. Truncated columns are normalized after recording
their discarded mass, and convergence is assessed at the level of stationary
observables as reported in Sec.~\ref{sec:numerics}.

Monte Carlo stationary samples are obtained by iterating
Eq.~\eqref{eq:affine_recursion} after burn-in. Transient trajectories sample the
exact number of events in each time interval from the Poisson law and then
sample the event types with probabilities $\lambda_k/\Lambda$. The computation
uses random seed 20260713, $3\times10^4$ transient trajectories, and $10^5$
stationary samples per selected $\eta$. All automated consistency tests reported
with this draft pass, including full resets, equal bath occupations, and a
weak-collision point at $\eta=0.9999$. Monte Carlo standard errors are computed
from independent trajectories; for ratios such as $g^{(2)}$, the code uses the
delta method for the joint sample moments $(\overline X,\overline{X^2})$.

\paragraph*{Code and data availability.}
The source code and numerical data supporting the figures, tables, Monte Carlo calculations, and Fock-space convergence checks are available at \url{https://github.com/iuriinosal/collision-bath-code}. The corresponding release (v1.0.0) is archived on Zenodo at \url{https://doi.org/10.5281/zenodo.22857152}.

\bibliography{references_v3.0.0}

\end{document}